\documentclass[superscriptaddress,twocolumn,pra,preprintnumbers,amsmath,amssymb,showpacs]{revtex4-1}
\usepackage{lipsum} 
\usepackage[latin9]{inputenc}
\usepackage{graphicx}
\usepackage{dcolumn}
\usepackage{bbold}
\usepackage{bm}
\usepackage[usenames]{xcolor}

\usepackage[colorlinks,bookmarks=false,citecolor=blue,linkcolor=blue,urlcolor=blue,hyperfootnotes=true]{hyperref}

\makeatletter
\newsavebox\myboxA
\newsavebox\myboxB
\newlength\mylenA

\newcommand*\xoverline[2][0.75]{%
    \sbox{\myboxA}{$\m@th#2$}%
    \setbox\myboxB\null
    \ht\myboxB=\ht\myboxA%
    \dp\myboxB=\dp\myboxA%
    \wd\myboxB=#1\wd\myboxA
    \sbox\myboxB{$\m@th\overline{\copy\myboxB}$}
    \setlength\mylenA{\the\wd\myboxA}
    \addtolength\mylenA{-\the\wd\myboxB}%
    \ifdim\wd\myboxB<\wd\myboxA%
       \rlap{\hskip 0.5\mylenA\usebox\myboxB}{\usebox\myboxA}%
    \else
        \hskip -0.5\mylenA\rlap{\usebox\myboxA}{\hskip 0.5\mylenA\usebox\myboxB}%
    \fi}
\makeatother

\begin{document}

\title{Thermal and thermoelectric transport coefficients for a two-dimensional SDW metal with weak disorder: A memory matrix calculation}

\author{Hermann Freire}

\affiliation{Instituto de Física, Universidade Federal de Goiás, 74.001-970, Goiânia-GO, Brazil}

\begin{abstract}
We calculate the thermal and thermoelectric transport coefficients for a two-dimensional spin-density-wave metal in the presence of weak disorder and under an external magnetic field using the Mori-Zwanzig memory matrix formalism. As a consequence, we obtain that although the Seebeck coefficient $S$ displays
a relatively conventional linear dependence on the temperature of the system, the Nernst coefficient $\nu$ exhibits a clear signature of bad-metallicity within the so-called strange metal regime described by the model. This result agrees qualitatively with experimental transport data obtained for many cuprate superconductors around optimal doping. 
\end{abstract}

\maketitle

\section{Introduction}

Calculating transport properties in non-Fermi-liquid (NFL) metallic phases is an extremely challenging task and, for this reason, it has become a research topic of paramount importance nowadays in the field of strongly correlated systems \cite{Zaanen}. Among the many properties
that turn out to be crucial in order to characterize those NFL quantum states, we cite in addition to the generally measured electrical conductivity $\sigma$, e.g., the thermal conductivity $\kappa$, the thermopower (Seebeck) coefficient $S$ and the Nernst coefficient $\nu$, to name only a few transport coefficients.
These are extremely useful physical quantities that provide important information regarding the character of the elementary excitations present in the system, the precise nature and possible transformations of the underlying Fermi surface and, ultimately, the microscopic mechanism of the NFL behavior displayed by many strongly correlated materials (see, e.g., Refs. \cite{Matsuda,Analytis} for recent examples involving iron-based superconductors). 

One well-known case of a NFL regime is the so-called strange metal phase that emerges around optimal doping in the cuprate superconductors (see, e.g., Refs. \cite{Ong,Ando,Raffy}). Today, there is a profusion of experimental transport data both
for this unconventional metallic phase and its accompanying mysterious pseudogap state that, despite more than three decades of intensive research performed on those materials, remain puzzling to this date \cite{Hussey}. For this reason, in order to focus our present discussion, we draw attention to some transport properties of the strange metal phase that are certainly noteworthy, i.e., the $T$-linear resistivity that extends well beyond the Ioffe-Regel limit at higher temperatures \cite{Ong,Freire4}, the bad-metal Nernst response of this NFL phase as a function of the temperature of the system \cite{Ong2,Cooper}, the sign-change of the prefactor of the $T$-linear Seebeck coefficient as a function of temperature and magnetic field and its possible connection to the reconstruction of the underlying Fermi surface inside the pseudogap phase in the underdoped regime \cite{Taillefer} and, last but not least, the behavior of the thermal conductivity together with the analysis of the fate of the so-called Wiedemann-Franz law that seems to fail in some situations for these materials \cite{Taillefer2}.

Technically speaking, the reason why is so difficult to calculate transport in NFL systems is based, on a fundamental level, on the fact that
the lack of coherent quasiparticle excitations at low energies in these regimes hampers a straightforward application of quantum Boltzmann equation for these systems \cite{Hartnoll2}.
Therefore, alternative computational methods that crucially do not rest on the assumption of the existence of well-defined quasiparticle excitations in order to calculate the transport properties of these systems are imperative. 
In this respect, we point out that the so-called Mori-Zwanzig memory matrix formalism \cite{Forster} turns out to be a very promising tool to describe NFL metallic regimes in an unbiased fashion.
This approach has been applied in recent years with good success to a wealth of fundamental condensed matter problems including one-dimensional models \cite{Rosch}, quantum spin liquids with a ``spinon" Fermi surface \cite{Freire300}, two-dimensional quantum critical metals \cite{Patel,Hartnoll3} and holographic quantum matter \cite{Zaanen,Hartnoll100}.

In this work, we perform a non-quasiparticle-based computation of the thermal and thermoelectric responses of a two-dimensional (2D) spin-density-wave (SDW) quantum critical metal in the presence of weak disorder and under the application of an external magnetic field using the memory matrix formalism.
The corresponding spin-fermion model was originally proposed by Abanov and Chubukov \cite{Chubukov1} as a possible low-energy effective theory to capture the essential physics displayed by the cuprate superconductors. Since then, it has been studied over the years by many different authors using a wide variety of theoretical techniques in the literature \cite{Metlitski,Efetov,SSLee,Wang,Hartnoll,Schattner}. The model assumes a central role of antiferromagnetic (AF) quantum criticality \cite{Sachdev} in these systems that stems from the assumed existence of an quantum critical point (QCP) buried under the superconducting dome. The AF fluctuations that it entails are expected to provide the ``pairing glue'' associated with the Cooper pair formation. However, it must be pointed out that the nature of this QCP in the cuprates is a hotly debated topic nowadays and many proposed candidates for the corresponding quantum phase transition in these systems include (but are not exhausted by) antiferromagnetic SDW order \cite{Chubukov1,Metlitski, Efetov}, pair-density-wave (PDW) \cite{Wang2,Freire2,Hamidian}, fractionalized Fermi-liquid (FL*) order \cite{Berg}, loop-current phases \cite{Varma,Freire1}, etc. Sorting out all the consequences of these many different proposed orders that could be present in the cuprate superconductors is an important focus of research in the field.

The perspective that we shall take from the outset in this work is that antiferromagnetic fluctuations is an important ingredient to describe the many unconventional properties of the strange metal phase of the cuprates around optimal doping. We will test this hypothesis by showing specifically that the spin-fermion model indeed captures qualitatively some essential aspects of the physics of these compounds from the point of view of transport.
For this reason, the outline of the present work will be as follows: First, we define the spin-fermion model that aims to provide a description of the strange metal phase of the cuprate superconductors around optimal doping. Then, we briefly explain 
the memory matrix methodology in order to calculate all transport coefficients of the model using this framework, which, as will become clear, lies beyond the quasiparticle paradigm. Next, we present our main results obtained in the present work by using this method
and show that the corresponding transport coefficients agree qualitatively with the experimental observation in these materials. Finally,
we present our general conclusions and the outlook regarding the present study.

\section{Spin-fermion model}

To initiate our discussion, we define the 2D model that describes a SDW quantum critical metal around optimal doping \cite{Chubukov1} with the subsequent addition of a higher-order effective composite interaction \cite{Hartnoll,Schmalian}. The effect of this composite interaction is to provide an efficient coupling to the whole Fermi surface of the normal state of the cuprates, which can be precisely measured by means of, e.g., ARPES experiments \cite{ZXShen}. As a result, the effective Lagrangian of the model is given by

\vspace{-0.3cm}

\begin{align}\label{1}
\mathcal{L}&=\sum_{\alpha}{\psi}^{\dagger}_{\alpha}(i\partial_{t}-v_F\partial_{x}-u'\partial_{y}^2)\psi_{\alpha}+\frac{1}{2}[(\partial_{t}\vec{\phi})^2-(\nabla\vec{\phi})^2]\nonumber\\
&-\frac{m_b^2}{2}\vec{\phi}^2-\frac{u}{4!}(\vec{\phi}^2)^2-\lambda\sum_{\alpha\alpha'}{\psi}^{\dagger}_{\alpha}(\vec{\phi}\cdot\vec{\sigma}_{\alpha\alpha'}){\psi}_{\alpha'}\nonumber\\
&-\lambda'\sum_{\alpha} {\psi}^{\dagger}_{\alpha}\psi_{\alpha}(\vec{\phi}\cdot\vec{\phi}),
\end{align}

\noindent where $\bar{\psi}_{\alpha}$ and ${\psi}_{\alpha}$ correspond to the (Grassmann) fermionic fields with spin $\alpha$, the quantity $v_F=(k_F/m)$ stands for the Fermi velocity of the system and $u'$ is the local curvature of the fermionic dispersion, $\vec{\phi}=(\phi_x,\phi_y, \phi_z)$ denotes the bosonic SDW collective field that describes the spin fluctuations in the model, $m_b$ is the corresponding ``mass'' of the bosonic field that vanishes at the QCP, $u$ is the bosonic self-interaction, and $\vec{\sigma}=(\sigma_x,\sigma_y, \sigma_z)$ represent conventionally the Pauli matrices. In addition, the parameter $\lambda$ refers to the fermion-boson coupling constant in the theory and $\lambda'$ is the composite interaction constant (to be explained below). Concerning the bosonic Green's function, we point out that we will restrict our present analysis to the situation around optimal doping (i.e., $m_b\approx 0$) and also to an intermediate temperature regime such that $T > E_s$ (where $E_s\sim \gamma$, with the parameter $\gamma \sim \lambda^2$ denoting the Landau damping constant of the model). In other words, we will depart here from the bosonic propagator at intermediate temperatures described by $\chi(q_0,\mathbf{q})=1/[q_{0}^{2}+\mathbf{q}^2+R(T)]$, where $q_0$ stands for the bosonic Matsubara frequency, $\mathbf{q}$ is the bosonic momentum peaked around the SDW ordering wavevector $(\pi,\pi)$ and $R(T)=4\ln^2[(\sqrt{5}+1)/2]T^2$ is
an infrared cutoff in the theory, which was computed previously in Ref. \cite{Chubukov3}. 

The model defined in Eq. (\ref{1}) emphasizes the role of special points at the Fermi surface of the system (the so-called ``hot spots''), which
are located precisely at the intersection of the underlying Fermi surface with the AF zone boundary connected by $(\pi,\pi)$. These are the points where the (bosonic) order-parameter field couples efficiently to the low-energy fermions of the system. 
Despite this statement, in view of the strong-coupling nature of SDW quantum criticality in two dimensions \cite{Metlitski}, the order-parameter fluctuations will couple not only to the ``hot spots'' explained above, but also to other regions of the underlying Fermi surface of the system. This was first pointed out by Hartnoll \emph{et al.} in Ref. \cite{Hartnoll}, where they have shown that such a higher-order effective composite interaction described by $\lambda'$ (which involves the low-energy fermion scattering off two spin fluctuations) emerges in the model and couples efficiently to the remaining parts of the Fermi surface. This makes these remaining regions of the underlying Fermi surface  at least ``lukewarm" (i.e., strongly renormalized), instead of simply ``cold'' (i.e., weakly renormalized) that was conventionally assumed in some pioneering transport calculations of a nearly antiferromagnetic metallic model within a Boltzmann equation approach in the literature \cite{Hlubina,Rosch2}. We mention here that the interplay of weak disorder with the above higher-order effective composite operator in other transport properties of the present spin-fermion model was also recently investigated, e.g., in the Refs. \cite{Schmalian,Freire100}.

By using the Noether's theorem and the fact that the system is initially invariant under continuous spacetime translations and global U(1) symmetry, we obtain that the total momentum, the energy and the electric current of the model are all conserved at the classical level. The corresponding canonical total momentum $\overrightarrow{\mathcal{P}}$ and the electric current $\overrightarrow{{J}}$ of the model are then given, respectively, by

\vspace{-0.3cm}

\begin{align}\label{3}
\overrightarrow{\mathcal{P}}&=\frac{1}{2}\sum_{\alpha}\int d^2 x \left[({\nabla{\psi}}^{\dagger}_{\alpha}\,\psi_{\alpha}-{\psi}^{\dagger}_{\alpha}{\nabla\psi}_{\alpha})+\nabla\vec{\phi}\cdot\partial_{t}\vec{\phi}\right],\\
\overrightarrow{{J}}&=-\frac{i}{m}\sum_{\sigma}\int d^2 x\, \nabla \bar{\psi}_{\sigma}\,\psi_{\sigma},
\end{align}

\noindent where the total momentum $\overrightarrow{\mathcal{P}}$ has both a fermionic contribution and a bosonic (drag) term, whereas the electric current $\overrightarrow{{J}}$ has naturally only the fermionic contribution (the bosonic spin fluctuations have no charge). 

The Hamiltonian density $h(\mathbf{x})$ of the spin-fermion model is naturally associated with the (0,0)-component of the corresponding energy-momentum tensor, which yields

\vspace{-0.3cm}

\begin{align}\label{2}
&h(\mathbf{x})=\sum_{\alpha}{\psi}^{\dagger}_{\alpha}(v_F\partial_{x}+u'\partial_{y}^2)\psi_{\alpha}+\frac{1}{2}[(\partial_{t}\vec{\phi})^2+(\nabla\vec{\phi})^2]\nonumber\\
&+\frac{u}{4!}(\vec{\phi}^2)^2+\lambda\sum_{\alpha\alpha'}{\psi}^{\dagger}_{\alpha}(\vec{\phi}\cdot\vec{\sigma}_{\alpha\alpha'}){\psi}_{\alpha'}+\lambda'\sum_{\alpha} {\psi}^{\dagger}_{\alpha}\psi_{\alpha}(\vec{\phi}\cdot\vec{\phi}).
\end{align}

\noindent If the energies are measured with respect to the chemical potential, the Hamiltonian density ${h}(\mathbf{x})$ of the model may be viewed as the heat density of the system. Therefore, by using the continuity equation for the heat flow $\dot{h}(\mathbf{x})+\nabla\cdot\mathbf{J}_Q=0$ (where the dot represents a time derivative), we can also formally obtain the thermal-current operator $\mathbf{J}_Q=({J}_Q^x,{J}_Q^y)$, whose components are given by

\vspace{-0.3cm}

\begin{align}\label{4b}
J_{Q}^{x}&=\frac{1}{2}\sum_{\alpha}\int d^2 x \left[v_F(\dot{{\psi}}^{\dagger}_{\alpha}\psi_{\alpha}-{\psi}^{\dagger}_{\alpha}\dot{\psi}_{\alpha})-\partial_x\vec{\phi}\cdot\partial_{t}\vec{\phi}\right],\\
J_{Q}^{y}&=\frac{1}{2}\sum_{\alpha}\int d^2 x \left[u'(\dot{{\psi}}^{\dagger}_{\alpha}\partial_y\psi_{\alpha}+\partial_y{\psi}^{\dagger}_{\alpha}\dot{\psi}_{\alpha})-\partial_y\vec{\phi}\cdot\partial_{t}\vec{\phi}\right].
\end{align}

\noindent In the above expressions, since both fermions and bosons in the model transport heat, the thermal-current operator must contain the two contributions, as expected.

\section{Memory matrix formalism}

As mentioned before, we will use here the Mori-Zwanzig memory matrix formalism that does not rely on the existence of well-defined quasiparticles in the model (for excellent, in-depth expositions on this method, see, e.g., Refs. \cite{Zaanen,Forster,Hartnoll100,Lucas}). This method draws inspiration from holographic methods applied to condensed matter problems and has been applied to many non-quasiparticle-based models with great success (see, e.g., Refs. \cite{Zaanen,Forster,Hartnoll100} and references therein).
Within the memory matrix approach, when an external magnetic field $\mathbf{B}$ is applied to the system, the ``generalized'' conductivities can be written as follows

\vspace{-0.5cm}

\begin{equation}\label{sigma}
\hat{\sigma}(\omega,T,B)=\frac{\hat{\chi}^R(T)}{(\hat{M}+\hat{N}-i\omega{\hat{\chi}^R(T)})[{\hat{\chi}^R}(T)]^{-1}},
\end{equation}

\noindent where $\hat{\chi}^R(T)$ denotes static retarded susceptibility matrices that represent the overlap of the currents of interest (i.e., electric, thermal or spin currents) with any almost-conserved operator (e.g., the physical total momentum $\mathbf{P}$ to be defined shortly) in the system. As an example of such a ``generalized" susceptibility, we have, e.g., $\chi_{{J_i}{P_j}}(i\omega,T)=\int_{0}^{1/T}d\tau e^{i\omega\tau}\langle T_{\tau}{J_i}^{\dagger}(\tau){P_{j}}(0)\rangle$, where the retarded susceptibility is naturally given by $\chi^R_{{J_i}{P_j}}(\omega)=\chi_{{J_i}{P_j}}(i\omega\rightarrow\omega +i0^{+})$. In addition, $\langle \ldots\rangle$ refers to the
grand-canonical ensemble average, $T_{\tau}$ corresponds to the time-ordering operator, and the volume $V$ of the system has been set, for simplicity, equal to unity.
As for the memory matrix $\hat{M}$, it can be computed by using the formally exact equation

\vspace{-0.3cm}

\begin{equation}\label{4}
\hat{M}_{P_i P_j}(T)=\int_{0}^{1/T}d\tau\left\langle \dot{{P_i}}^{\dagger}(0)Q\frac{i}{\omega-Q\hat{L}Q}Q\dot{{P_j}}(i\tau)\right\rangle,
\end{equation}

\noindent where the operator $\hat{L}$ is the so-called Liouville operator, which is defined as $\hat{L}\,\mathcal{O}=[H,\mathcal{O}]=-i\dot{\mathcal{O}}$, with $H$ corresponding to the Hamiltonian of the system and $\mathcal{O}$ is an arbitrary conserved or almost conserved operator in the system. The $Q$ is another operator that projects out of a space
spanned by all the conserved or nearly conserved operators in the system. In this way, the memory matrix encodes the relaxation mechanism of all the operators that are relaxed on long timescales in the present low-energy theory. Finally, the $\hat{N}$ represents a time-reversal symmetry breaking matrix associated, e.g., with the application of the external magnetic field, whose elements are given by $N_{P_i P_j}=\chi_{P_i\dot{P}_j}$. 

As was discussed previously, the spin-fermion model defined in Eq. (\ref{1}) conserves total momentum on a classical level. In this way, in order for the electric and thermal currents to decay in the present system, we must
specify a mechanism for relaxing the total momentum given by Eq. (\ref{3}). Coupling to phonons are of course a well-known source of momentum relaxation at higher temperatures (the so-called Bloch mechanism). However, we will focus here on an intermediate-to-low temperature regime where impurity effects are known to be important. For this reason, we will choose in this work both short-wavelength and long-wavelength disorder to
provide the microscopic mechanism that will effectively degrade the total momentum of the present system (this is the so-called Peierls mechanism \cite{Peierls}). As a result, in order to include disorder effects in the model, we must add the following terms to the Langrangian in Eq. (\ref{1}), i.e.,

\vspace{-0.3cm}

\begin{eqnarray}\label{2}
\mathcal{L}_{imp}&=&\sum_{\sigma}V(\vec{r})\bar{\psi}_{\sigma}(\vec{r})\psi_{\sigma}(\vec{r})+\sum_{\sigma}m(\vec{r})\vec{\phi}(\vec{r})\cdot\vec{\phi}(\vec{r}),\nonumber\\
\end{eqnarray}

\noindent which should naturally obey the standard (Gaussian) disorder averages: $\langle\langle V(\vec{r}) \rangle\rangle=\langle\langle m(\vec{r}) \rangle\rangle=0$,
$\langle\langle V(\vec{r})V(\vec{r'}) \rangle\rangle=V_0^2\delta^2(\vec{r}-\vec{r'})$, and $\langle\langle m(\vec{r})m(\vec{r'}) \rangle\rangle=m_0^2\delta^2(\vec{r}-\vec{r'})$, where $V_0$ is a random potential
for the fermion field and the parameter $m_0$ is a random mass term for the boson field. 

Due to the addition of weak disorder to the Lagrangian of the model, the canonical total momentum is not
conserved any longer in view of the breaking of the continuous translation symmetry in the system. Hence, the equation of motion that describes the time-evolution of the canonical momentum $\overrightarrow{\mathcal{P}}$ becomes

\vspace{-0.3cm}

\begin{eqnarray}\label{6}
i\dot{\overrightarrow{\mathcal{P}}}&=&\int \frac{d^2\mathbf{q}}{(2\pi)^2}\int \frac{d^2\mathbf{k}}{(2\pi)^2}\mathbf{k}\bigg[V(\mathbf{k})\sum_{\sigma}\bar{\psi}_{\sigma}(\mathbf{k}+\mathbf{q})\psi_{\sigma}(\mathbf{k})\nonumber\\
&+&m(\mathbf{k})\phi(\mathbf{q})\phi(\mathbf{-q-k})\bigg].
\end{eqnarray}

\noindent As an external magnetic field $\mathbf{B}$ is applied to the system, we have to perform the following substitution in the Lagrangian of the system, i.e., $\mathcal{L}\rightarrow \mathcal{L}+\int d^2 x \,\mathbf{j}(\mathbf{x})\cdot\mathbf{A}(\mathbf{x})$, where $\mathbf{j}(\mathbf{x})$ is the current density and $\mathbf{A}(\mathbf{x})$ is the vector potential such that $\mathbf{B}=\nabla \times \mathbf{A}$. Hence, the physical momentum $\mathbf{P}$ becomes different from the canonical momentum $\overrightarrow{\mathcal{P}}$ in the following way

\vspace{-0.3cm}

\begin{eqnarray}
\mathbf{P}=\overrightarrow{\mathcal{P}}-\int d^2 x \, \rho(\mathbf{x}) \mathbf{A}(\mathbf{x}),
\end{eqnarray}

\noindent where $\rho(\mathbf{x})$ is the charge density that obeys the continuity equation: $\partial \rho/\partial t + \nabla\cdot\mathbf{j}(\mathbf{x})=0$. In
the present case, we point out that the only nearly conserved operator in the model that will dominate the transport properties will be given by the physical total momentum mode denoted by $\mathbf{P}=(P_x,P_y)$.

\section{Results}

By calculating the overlap of the $x$-component of the heat current with the total momentum of the system given by the static retarded susceptibility $\chi_{P_x,J_Q^x}(T)\equiv\langle P_x | J_Q^x \rangle$, we obtain the result

\vspace{-0.3cm}

\begin{align}\label{4a}
&\chi_{P_x,J_Q^x}(T) = -T\sum_{k_0}\bigg[\int_{\mathbf{k}}\frac{v_F k_x(k_0+q_0/2)}{(ik_0+iq_0-\bar{\varepsilon}_{\mathbf{k+q}})(ik_0-\bar{\varepsilon}_{\mathbf{k}})}\nonumber\\
&+\int_{\mathbf{k}}\frac{k_x^2(k_0+q_0)^2}{[(k_0+q_0)^2+(\mathbf{k}+\mathbf{q})^2+R(T)][k_0^2+\mathbf{k}^2+R(T)]}\bigg]\nonumber\\
&\approx \frac{i\Lambda_{\parallel}}{6 v_F} T^2+1.56\,T^3,
\end{align}

\noindent where the limit $q_0\rightarrow 0$ must be taken in the calculation, and $\Lambda_{\parallel}$ is an ultraviolet cutoff that must be imposed in the momentum component parallel to the Fermi surface. We note that the $T^2$-term in the result displayed above corresponds to the contribution of the fermions to the overlap of the heat current with the total momentum, while the $T^3$-contribution refers to the bosonic contribution. Therefore, we can conclude from the above ``generalized'' susceptibility that the bosonic contribution turns out to be subleading at lower temperatures with respect to the fermionic counterpart.
In this way, we may neglect the bosonic contribution obtained in Eq. (\ref{4a}) to leading order in temperature, and keep only the fermionic contribution in the calculation, i.e., $\chi_{P_x,J_Q^x}(T)\sim T^2$ within an intermediate-to-low temperature regime. Analogously, we can also calculate the corresponding overlap of the the $y$-component of the heat current with the momentum of the system given by the static retarded susceptibility $\chi_{P_y,J_Q^y}(T)\equiv\langle P_y | J_Q^y \rangle$. In this case, this quantity yields the following scaling form $\chi_{P_y,J_Q^y}(T) \sim\,T^0$.

The thermopower (Seebeck) coefficient is conventionally defined as $S=(\alpha_{xx}/\sigma_{xx})$, where $\alpha_{xx}$ is the longitudinal Peltier coefficient and $\sigma_{xx}$ is the dc longitudinal conductivity. As for the Nernst coefficient $\nu$, it is given by the standard relation

\vspace{-0.3cm}

\begin{equation}\label{Nernst}
\nu=\frac{1}{B}\left[\frac{\alpha_{xy}}{\sigma_{xx}}-S \tan(\theta_H)\right],
\end{equation}

\noindent where $\theta_H$ is the Hall angle, i.e., $\tan(\theta_H)=\sigma_{xy}/\sigma_{xx}$. Within the memory matrix formalism,
these quantities are defined in terms of the ``generalized'' dc conductivities
${\sigma}=\hat{\sigma}(\omega\rightarrow 0)=\hat{\chi}^R(T)(\hat{M}+\hat{N})^{-1}\hat{\chi}^R(T)$.
In this way, the electrical conductivity $\sigma_{ij}$ and the thermal conductivity at zero electric field $\kappa_{ij}$ are defined, respectively, by
$\sigma_{ij}=\sigma_{J_i J_j}$, $\kappa_{ij}=\frac{\sigma_{J_{Q}^i J_{Q}^j}}{T}$, whereas the Peltier coefficients are in turn defined by
$\alpha_{ij}=\frac{\sigma_{J_i J_{Q}^j}}{T}$, for $i=x,y$ representing the components of the corresponding currents.

Now, we proceed to calculate the memory matrix of the system. Using the gauge choices $\mathbf{A}=B(-y,0)$ and $\mathbf{A}=B(0,x)$, we obtain that the time-reversal symmetry breaking matrix finally evaluates to $N_{P_x P_y}=-N_{P_y P_x}=-B\chi_{J_xP_x}$ and $N_{P_x P_x}=N_{P_y P_y}=0$.
Therefore, to lowest order in the magnetic field $B$, we get

\vspace{-0.3cm}

\begin{equation}
(\hat{M}+\hat{N})^{-1}\approx \frac{1}{\det(\hat{M}+\hat{N})}\left( \begin{array}{ccc}
M_{P_yP_y} & B\chi_{J_xP_x} \\
-B\chi_{J_xP_x} & M_{P_xP_x} \end{array} \right).
\end{equation}

\noindent As a result, we obtain that the longitudinal electrical conductivity ${\sigma}_{xx}$, the thermal conductivity ${\kappa}_{xx}$, the transverse thermal conductivity  ${\kappa}_{xy}$, and the Peltier coefficients  ${\alpha}_{xx}$ and ${\alpha}_{xy}$ are given, respectively, by

\vspace{-0.3cm}

\begin{eqnarray}
{\sigma}_{xx}(T,B)&=&\frac{\chi^2_{J_x P_x} M_{P_x P_x}}{(M_{P_x P_x}^2+B^2 \chi^2_{J_x P_x})},\\
{\kappa}_{xx}(T,B)&=&\frac{1}{T}\frac{\chi^2_{J_Q^x P_x} M_{P_x P_x}}{(M_{P_x P_x}^2+B^2 \chi^2_{J_x P_x})},\\
{\kappa}_{xy}(T,B)&=&\frac{1}{T}\frac{B\chi^2_{J_Q^x P_x}\chi_{P_y J_Q^y}}{(M_{P_x P_x}^2+B^2 \chi^2_{J_x P_x})},\\
{\alpha}_{xx}(T,B)&=&\frac{1}{T}\frac{\chi_{J_x P_x} M_{P_x P_x}\chi_{P_x J_Q^x}}{(M_{P_x P_x}^2+B^2 \chi^2_{J_x P_x})},\\
{\alpha}_{xy}(T,B)&=&\frac{1}{T}\frac{B\chi^2_{J_x P_x}\chi_{P_y J_Q^y}}{(M_{P_x P_x}^2+B^2 \chi^2_{J_x P_x})}.
\end{eqnarray}

\noindent From the above equations, one can clearly see that it is essential to analyze also the overlap of the total momentum with the electric current described by static retarded susceptibility $\chi_{J_x P_x}(T)\equiv\langle J_x | P_x \rangle$. 
Indeed, this quantity yields, to leading order, the following result

\vspace{-0.3cm}

\begin{eqnarray}\label{9}
\chi_{J_x P_x}(T)&=&\frac{1}{m}\int \frac{d^2\mathbf{k}}{(2\pi)^2} k_x^2 \frac{[n_F(\bar{\varepsilon}_{\mathbf{k}})-n_F(\bar{\varepsilon}_{\mathbf{k+q}})]}{(\bar{\varepsilon}_{\mathbf{k}}-\bar{\varepsilon}_{\mathbf{k+q}})}
\nonumber\\
&\approx&\frac{m}{\pi} \mu(T=0)+O(e^{-\beta \mu}),
\end{eqnarray}

\noindent where $\mu$ is the chemical potential that controls the doping in the model, $n_F(\varepsilon)=1/(e^{\beta\varepsilon}+1)$ is the Fermi-Dirac distribution and $\beta=1/T$ is the inverse temperature. Therefore, the above susceptibility turns out to be temperature-independent. This result should be contrasted with Eq. (\ref{4a}), where the overlap of the total momentum with the thermal current is temperature-dependent.

The next step in our transport theory is to perform a perturbative calculation of the memory matrix for the present model.
As a first approximation, we will assume that the parameters $\lambda$, $\lambda'$, $V_0$, and $m_0$ are effectively small in the present theory. Since Eq.  (\ref{6}) turns out to be of order linear in both $V_0$ and $m_0$, the
leading contribution to the memory matrix will in fact be quadratic in those parameters. Similarly, the most important contribution to the Liouville operator defined after Eq.  (\ref{4}) will be given by its non-interacting value (i.e., $\hat{L}\approx \hat{L}_0$) and the dominant contribution in the grand-canonical ensemble average should be expressed also in terms of the non-interacting Hamiltonian of the system. The corresponding Feynman diagrams associated with this calculation are depicted in Fig. 1. In this way, it can be readily demonstrated that the most important contribution to the memory matrix reads as follows:
$M_{P_x P_x}(\omega\rightarrow 0,T)=\lim_{\omega\rightarrow 0}\frac{\text{Im}\,G^{R}_{\dot{P_x}\dot{P_x}}(\omega,T)}{\omega},$
where $G^{R}_{\dot{P_x}\dot{P_x}}(\omega,T)=\langle\dot{P_x}(\omega)\dot{P_x}(-\omega)\rangle$ refers to a retarded Green's function associated with the operators $\dot{P_x}(\omega)$ and $\dot{P_x}(-\omega)$ at finite temperatures. Since the bosonic self-interaction $u$ and the local curvature in the fermionic dispersion $u'$ in Eq. (\ref{1}) are irrelevant in the renormalization group (RG) sense \cite{SSLee,Freire100}, we will neglect, for simplicity, both terms in what follows. 

From Fig. 1, one can see that the memory matrix can be written as
$M_{P_xP_x}(T)=\sum_{i=0}^{5}M^{(i)}(T)+\dots$,
where the index $i$ denotes, respectively, each Feynman diagram shown in this figure. The computation of these specific diagrams for the spin-fermion model has been performed in detail by the present author elsewhere \cite{Freire100} and, for this reason, we will not repeat all the technical details regarding this calculation here. 
By computing the Feynman diagram with label (0) in Fig. 1, we obtain that this contribution naturally evaluates to $M^{(0)}=-\sum_{i,j,i\neq j}Q^2_{ij} V_0^2  \Lambda^2/(4\pi^3 |{\vec{v}_{i}}\times{\vec{v}_{j}}|)$,
where $\Lambda$ is an ultraviolet cutoff that must be defined in the integration
over all the energies in the theory and the quantity $Q_{ij}$ is the (large) momentum transfer connecting the so called ``hot spots'' represented by arbitrary indices $i,j$ with Fermi velocities denoted by $v_i$ and $v_j$ in the theory. Naturally, this latter result turns out to be a temperature-independent contribution to the memory matrix of the present model.

\begin{figure}[t]
\includegraphics[width=3.23in]{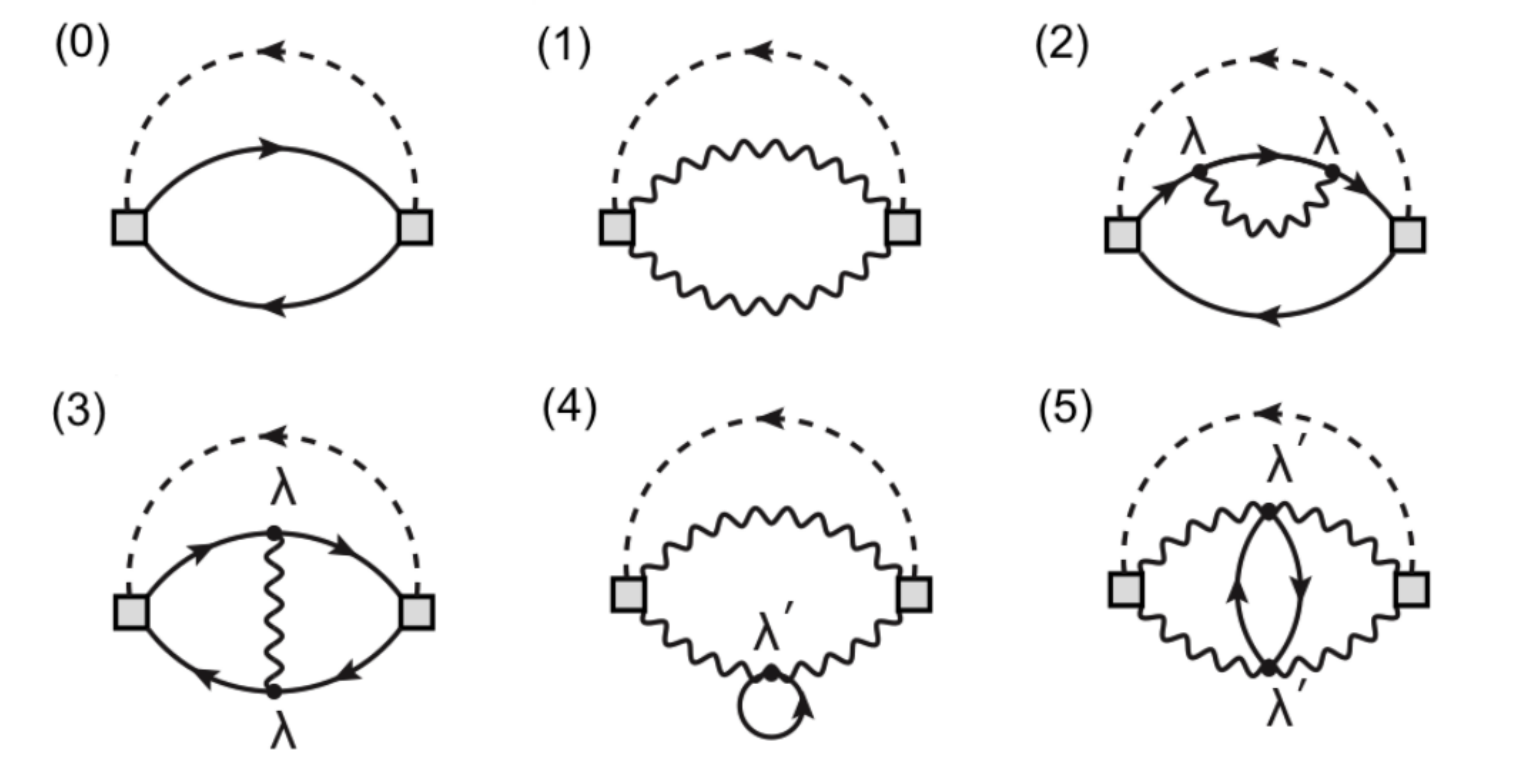}
\caption{Some Feynman diagrams associated with the computation of $G^{R}_{\dot{P_x}\dot{P_x}}(\omega,T)$ in the present model. The solid lines correspond to the fermionic Green's function, while the wave lines correspond to the propagator of the bosons. The dashed lines refer to the impurity lines that carry only internal momentum and
external bosonic energy $\omega$.}\label{Feynman_diagrams}
\end{figure}

Next, we calculate the contribution with label (1) in Fig. 1. By using the Feynman rules for this model, this diagram yields the following result: $M^{(1)}(T)\approx(0.011 m_{0}^2)T^2$. From this expression, we conclude that the prefactor of such a Fermi-liquid-like result turns out to be isotropic and
doping independent. This term effectively arises as a result of the long-wavelength disorder considered in the present system. Moreover, the term denoted by the label (2) in Fig. 1 is a scattering process involving a self-energy correction to the fermionic propagator in the model that naturally yields a vanishing result. Indeed, it can be analytically demonstrated that all Feynman diagrams up to second order in $m_0$, $V_0$, $\lambda$, and $\lambda'$ for the memory matrix with self-energy insertions in both fermionic and bosonic propagators [see also, e.g., diagram (4)] evaluates to zero in the present formalism. 

Then, we proceed to calculate the diagram with label (3) in Fig. 1 that emerges as a result of the interplay of inter-hot-spot scattering and short-wavelength disorder in the spin-fermion model. The corresponding Feynman diagram naturally yields the following result:
$M^{(3)}(T)\approx \sum_{i,j,i\neq j}\sum_{\alpha,\beta}\left(\frac{0.001 V_{0}^2 \lambda^{2} Q^{2}_{ij}
}{ | \vec{v}_{i\alpha}\times \vec{v}_{i\beta}| | \vec{v}_{j\alpha}\times \vec{v}_{j\beta}|}\right)T$,
which clearly depends on the band structure of the model through the Fermi velocities $v_i$ and $v_j$ (i.e., it is doping-dependent). Such a $T$-linear contribution to the memory matrix leads to the well-known (non-Fermi-liquid-like) linear resistivity observed in the strange metal phase that emerges in the cuprate superconductors around optimal doping. This robust transport property is indeed a hallmark of such a NFL phase.
Finally, we compute the contribution with label (5) in Fig. 1, which is related to the effect of the composite operator in the transport properties of the model. In this case, we get
$M^{(5)}\approx\sum_{i,j,i\neq j}\left(\frac{0.96 m_{0}^2 \lambda'^{2}}{256\pi^2|\vec{v}_{i}\times\vec{v}_{j}|}\right)\Lambda^2$. Since this latter result is also temperature-independent, it will contribute as well to the residual resistivity $\rho_0$ of the system.

Collecting all the contributions obtained above, we conclude that the memory matrix of the present model may be schematically written as $M_{P_x P_x}(T)\sim A_1+A_2 T+A_3 T^2$, where $A_1$, $A_2$ and $A_3$ are temperature-independent prefactors. In what follows, we will neglect the residual contribution to the memory matrix described by the coefficient $A_1$ that turns out to be important only at very low temperatures in the system  (in this respect, we note that in such a low-temperature regime our present theory does not apply any longer). Therefore, the $T$-linear contribution will dominate the memory matrix calculation for this regime.
As a consequence of this, we obtain that the scaling forms concerning the temperature dependence of the following physical quantities, i.e., the longitudinal electrical conductivity ${\sigma}_{xx}$, the thermal conductivity ${\kappa}_{xx}$, the transverse thermal conductivity  ${\kappa}_{xy}$, and the Peltier coefficients  ${\alpha}_{xx}$ and ${\alpha}_{xy}$ of the model at optimal doping are given, to leading order, by the following expressions

\vspace{-0.3cm}

\begin{eqnarray}
{\sigma}_{xx}(T,B)&\sim&\frac{1}{T},\label{11}\\
{\kappa}_{xx}(T,B)&\sim&T^{2},\label{12}\\
{\kappa}_{xy}(T,B)&\sim&B\, T,\label{13}\\
{\alpha}_{xx}(T,B)&\sim&T^{0},\label{14}\\
{\alpha}_{xy}(T,B)&\sim&B\,T^{-3},\label{15}
\end{eqnarray}

\noindent which are valid of course only at intermediate temperatures in the present model. In a previous paper,
we proposed a scenario \cite{Freire200} in order to explain the apparent ``separation of lifetimes'' in the strange metal phase of the cuprates regarding the tangent of the Hall angle $\theta_H$, which was defined previously in this work. As a consequence, we showed that this quantity is described by $\tan\theta_H\sim C_1/T+C_2/T^2$ (where $C_2\gg C_1)$ around optimal doping. This arises due to the emergent particle-hole symmetry close to the ``hot spots" of the theory, which emerges as a result of
the renormalization of the underlying Fermi surface of the system. This proposed scenario agrees well with many theoretical RG calculations available in the literature applied to this model \cite{Chubukov1,Metlitski,SSLee,Freire2} and also with experimental observation \cite{Ando,Raffy}. 

In addition to this, from Eqs. (\ref{11})-(\ref{15}),
we are finally able to obtain the scaling forms of the Seebeck coefficient $S$ and Nernst response $\nu$ in the strange metal phase of the model. Using the previous definitions for both physical quantities, we obtain, to leading order, that

\vspace{-0.3cm}

\begin{equation}
S\sim -C_3\,T\hspace{1cm}\text{and}\hspace{1cm} \nu\sim \frac{C_4}{T}+\frac{C_5}{T^2},
\end{equation}

\noindent where $C_3$, $C_4$ and $C_5$ are temperature-independent prefactors. From the above expressions, we observe that although the temperature-dependence of the Seebeck coefficient turns out to be relatively conventional (i.e., ``Fermi-liquid"-like) in this NFL phase, the Nernst response associated with this quantum state clearly indicates bad-metallicity \cite{Behnia}. Remarkably, the above scaling forms agree qualitatively with many experimental data available for the cuprate compounds (see, e.g., Refs. \cite{Ong2,Cooper,Kim,Taillefer}). Notwithstanding this statement, we point out that data over a larger temperature range are probably necessary in order to establish precisely the temperature dependence of those transport coefficients in these materials.

\section{Conclusions}

To sum up, we have calculated several transport coefficients associated with the two-dimensional spin-fermion model in the presence of weak disorder by using the memory matrix approach. As a consequence, we have obtained several scaling forms regarding the temperature dependence of these transport quantities for this model around optimal doping at intermediate temperatures. These results provide clear-cut predictions, which can eventually be either confirmed or ruled out experimentally inside the strange metal phase of the cuprate superconductors. Further extensions of the present study would consist of describing the transport coefficients for the unconventional superconductivity that emerges at lower temperatures from the strange metal phase described here. This would imply of course generalizing the present formalism to the situation, in which spontaneously broken gauge symmetries play a central role in describing the transport properties of the model.

\acknowledgments
I would like to thank the Brazilian agency CNPq under grant No. 405584/2016-4 for financial support.

\end{document}